\documentclass[aps,prd,preprint,superscriptaddress,tightenlines,nofootinbib]{revtex4}

\usepackage{epsfig}
\usepackage{graphicx}
\usepackage{dcolumn}
\usepackage{bm}

\newcommand{\etal}{{\it et al.}}

\begin{document}

\preprint{\tighten\vbox{\hbox{\hfil CLNS 04/1896}
                        \hbox{\hfil CLEO 04-15}}}

\title{\LARGE Measuring ${\cal{B}}(D^+\to\mu^+\nu)$ and the
Pseudoscalar Decay Constant $f_{D^+}$}

\author{G.~Bonvicini}
\author{D.~Cinabro}
\author{M.~Dubrovin}
\affiliation{Wayne State University, Detroit, Michigan 48202}
\author{A.~Bornheim}
\author{S.~P.~Pappas}
\author{A.~J.~Weinstein}
\affiliation{California Institute of Technology, Pasadena,
California 91125}
\author{J.~L.~Rosner}
\affiliation{Enrico Fermi Institute, University of Chicago,
Chicago, Illinois 60637}
\author{R.~A.~Briere}
\author{G.~P.~Chen}
\author{T.~Ferguson}
\author{G.~Tatishvili}
\author{H.~Vogel}
\author{M.~E.~Watkins}
\affiliation{Carnegie Mellon University, Pittsburgh, Pennsylvania
15213}
% ...
\collaboration{CLEO Collaboration} 
\noaffiliation


\date{November 11, 2004}

\begin{abstract}
In 60 pb$^{-1}$ of data taken on the $\psi(3770)$ resonance with
the CLEO-c detector, we find 8 $D^+\to\mu^+\nu$ decay candidates
that are mostly signal, containing only 1 estimated background.
Using this statistically compelling sample, we measure a value of
${\cal{B}}(D^+\to\mu^+\nu)=(3.5\pm 1.4 \pm 0.6)\times 10^{-4}$,
and determine $f_{D^+}=(202\pm 41\pm 17)$ MeV.
\end{abstract}


\maketitle
\tighten

\newpage
\section{Introduction}
Measuring purely leptonic decays of heavy mesons allows the
determination of  meson decay constants, which connect measured quantities,
such as the $B\bar{B}$  mixing ratio, to CKM matrix elements.
Currently, it is not possible to determine $f_B$ experimentally from
leptonic $B$ decays, so theoretical calculations of $f_B$ must be used.
The most promising of these calculations involves lattice
QCD \cite{Davies,Lat:Milc,Lat:UKQCD},
though there are other methods \cite{Equations,Chiral,Sumrules,Quarkmodel,Isospin}.

Measurements of pseudoscalar decay
constants such as $f_{D^+}$ provide checks on these calculations and
help discriminate among different models.

\begin{figure}[htbp]
\centerline{ \epsfxsize=3.0in \epsffile{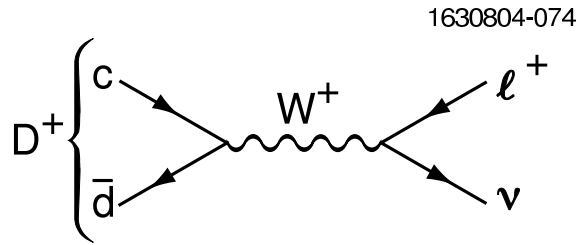} }
\caption{The decay diagram for $D^+\to \mu^+\nu$.} \label{Dptomunu}
\end{figure}

The decay diagram for $D^+\to \mu^+\nu$ is shown in
Fig.~\ref{Dptomunu}. The decay rate is given by \cite{Formula1}
\begin{equation}
\Gamma(D^+\to \ell^+\nu) = {{G_F^2}\over 8\pi}f_{D^+}^2m_{\ell}^2M_{D^+}
\left(1-{m_{\ell}^2\over M_{D^+}^2}\right)^2 \left|V_{cd}\right|^2~~~,
\label{eq:equ_rate}
\end{equation}
where $M_{D^+}$ is the $D^+$ mass, $m_{\ell}$ is the mass of the final
state lepton, $V_{cd}$ is a CKM matrix element equal to 0.224
\cite{PDG}, and $G_F$ is the Fermi coupling constant. Various
theoretical predictions of  $f_{D^+}$ range from 190 MeV to 350
MeV \cite{Davies,Lat:Milc,Lat:UKQCD,Equations,Chiral,Sumrules,Quarkmodel,Isospin}. Because of helicity suppression, the electron
mode $D^+ \to e^+\nu$ has a very small rate in the Standard Model \cite{Akeroyd}. The relative widths
are $2.64:1:2.3\times 10^{-5}$ for the $\tau^+ \nu$, $\mu^+ \nu$
and $e^+ \nu$ final states, respectively. Unfortunately the mode
with the largest branching fraction,
 $\tau^+\nu$, has at least
 two neutrinos in the final state and is difficult to detect.

\section{The CLEO-c Detector}

The CLEO-c detector is equipped to measure the momenta and
direction of charged particles, identify charged hadrons, detect
photons, and determine with good precision their directions and
energies. Muons above 1.1 GeV can also be identified. The detector
is almost cylindrically symmetric with everything but the muon
detector inside a superconducting magnet coil run at a current
that produces an almost uniform 1.0 T field. The detector consists
of a six-layer wire drift chamber at small radius that is low
mass, suitable for these relatively low energies. It is followed
by a 47-layer drift chamber; both chambers use a gas mixture of
60\% Helium and 40\% Propane. These two devices measure charged
track three-momenta with excellent accuracy. The drift chamber
also measures energy loss, dE/dx, that is used to identify charged
tracks below about 0.7 GeV \cite{CLEODR}. After the drift chamber there is a
Ring Imaging Cherenkov Detector (RICH) \cite{RICH}, that
identifies charged particles over most of their momentum range.
The RICH is surrounded by a Thallium doped CsI crystal array
consisting of about 8000 tapered crystals, 30 cm long and about 5x5
cm$^2$ at the front \cite{CLEOD}.

\section{Data Sample and Signal Selection}

In this study we use 60 pb$^{-1}$ of CLEO-c data produced in $e^+e^-$ collisions and recorded at the
$\psi''$ resonance (3.770 GeV).  At this energy, the events consist of a mixture
of pure $D^+D^-$, $D^o\overline{D}^o$ and three-flavor continuum events, resulting from the production of $u\overline{u}$, $d\overline{d}$ or $s\overline{s}$ quark pairs. There also
may be small amounts of $\tau^+\tau^-$ pairs and two-photon
events.

We examine all the recorded events and retain
those containing at least one charged $D$ candidate in the modes listed in
Table~\ref{tab:Drecon}. The selection criteria are described in detail in what
follows. We then use this sample to look for cases where we have only a single muon candidate whose four-momentum is consistent with a two-body $D$ decay into a muon and a neutrino and no other charged tracks or excess neutral energy are present.


All acceptable track candidates must have a helical trajectory that
approaches the event origin within a distance of 5 mm in the
azimuthal projection and 5 cm in the polar view, where the
azimuthal projection is in the bend view of the solenoidal magnet.
Each track must possess at least 50\% of the hits expected to be on a
track, and it must be within the fiducial volume of the drift chambers, $|\cos\theta|<0.93$, where $\theta$ is the polar angle with respect to the beam direction.


We use both charged particle ionization loss in the drift chamber
(dE/dx) and RICH information to identify kaons and pions used to
fully reconstruct $D$ mesons. The RICH is used for momenta larger
than 0.55 GeV. Information on the angle of detected Cherenkov
photons is translated into a likelihood of a given photon being due
to a particular particle. Contributions from all photons
associated with a particular track are then summed to form an
overall likelihood denoted as ${\cal L}_i$ for each particle
hypothesis. To differentiate between pion and kaon candidates, we
use the difference: $-2\log({\cal L_{\pi}})+2\log({\cal L}_K$).
Usually this cut is set at zero except for muon candidates where
the difference $-2\log({\cal L_{\mu}})+2\log({\cal L}_K$) is
required to be less than 10, to ensure a high, well understood efficiency. To
utilize the dE/dx information we calculate  $\sigma_{\pi}$ as the
difference between the expected ionization loss for a pion and the
measured loss divided by the measurement error.  Similarly,
$\sigma_{K}$ is defined  in the same manner using the expected
ionization for a  kaon .

We use both the RICH and dE/dx information for $D^-$ meson tag
candidate tracks in the following manner: (a) If neither the RICH
nor dE/dx information is available, then the track is accepted as both a pion and a kaon candidate. (b)
If dE/dx is available and RICH is not then we insist that pion
candidates have $PID_{dE}\equiv\sigma_{\pi}^2-\sigma_{K}^2 <0$, and kaon
candidates have $PID_{dE}> 0.$ (c) If RICH information is available
and dE/dx is not available, then we require that
$PID_{RICH}\equiv -2\log({\cal L}_{\pi})+2\log({\cal L}_K)<0$ for pions
and $PID_{RICH}>0$ for kaons. (d) If both dE/dx and RICH
information are available, we require that $(PID_{dE}+PID_{RICH})
<0$ for pions and $(PID_{dE}+PID_{RICH})>0$ for kaons.


We reconstruct $\pi^o$'s by first selecting photon candidates from
energy deposits in the crystals not matched to charged tracks that
have deposition patterns consistent with that expected for
electromagnetic showers. Pairs of photon candidates are kinematically fit to the
known $\pi^o$ mass.  We require the pull, the difference between
the raw and fit mass normalized by its uncertainty, to be less
than three for acceptable $\pi^o$ candidates.

 $K_S$ candidates are formed from a pair of charged pions  which are constrained to
 come from a single vertex. We also require that the invariant mass of the two
 pions be within 4.5 times the width of the $K_S$ mass peak, which has an r.m.s. width of 4 MeV.

\section{Reconstruction of Charged ${\boldmath D}$ Tagging Modes}

Tagging modes are fully reconstructed by first evaluating the
difference in the energy, $\Delta E$, of the decay products with the beam
energy. We then require the absolute value of this
difference to be within 20 MeV of zero, approximately twice the
r.m.s. width, and then look at the reconstructed $D^-$
beam-constrained mass defined as
\begin{equation}
m_D=\sqrt{E_{beam}^2-(\sum_i\overrightarrow{p}_{\!i})^2},
\end{equation}
where $i$ runs over all the final state particles.
The beam-constrained mass has better resolution then merely calculating
the invariant mass of the decay products since the beam has a small
energy spread. Besides using $D^-$ tags and searching for $D^+\to\mu+\nu$,
we also use the charge-conjugate $D^+$ tags and search for $D^-\to
\mu^-\overline{\nu}_{\mu}$; in the rest of this paper we will not
mention the charge-conjugate modes explicitly, but they are always
used.

\begin{figure}[htbp]
\centerline{ \epsfxsize=6.0in \epsffile{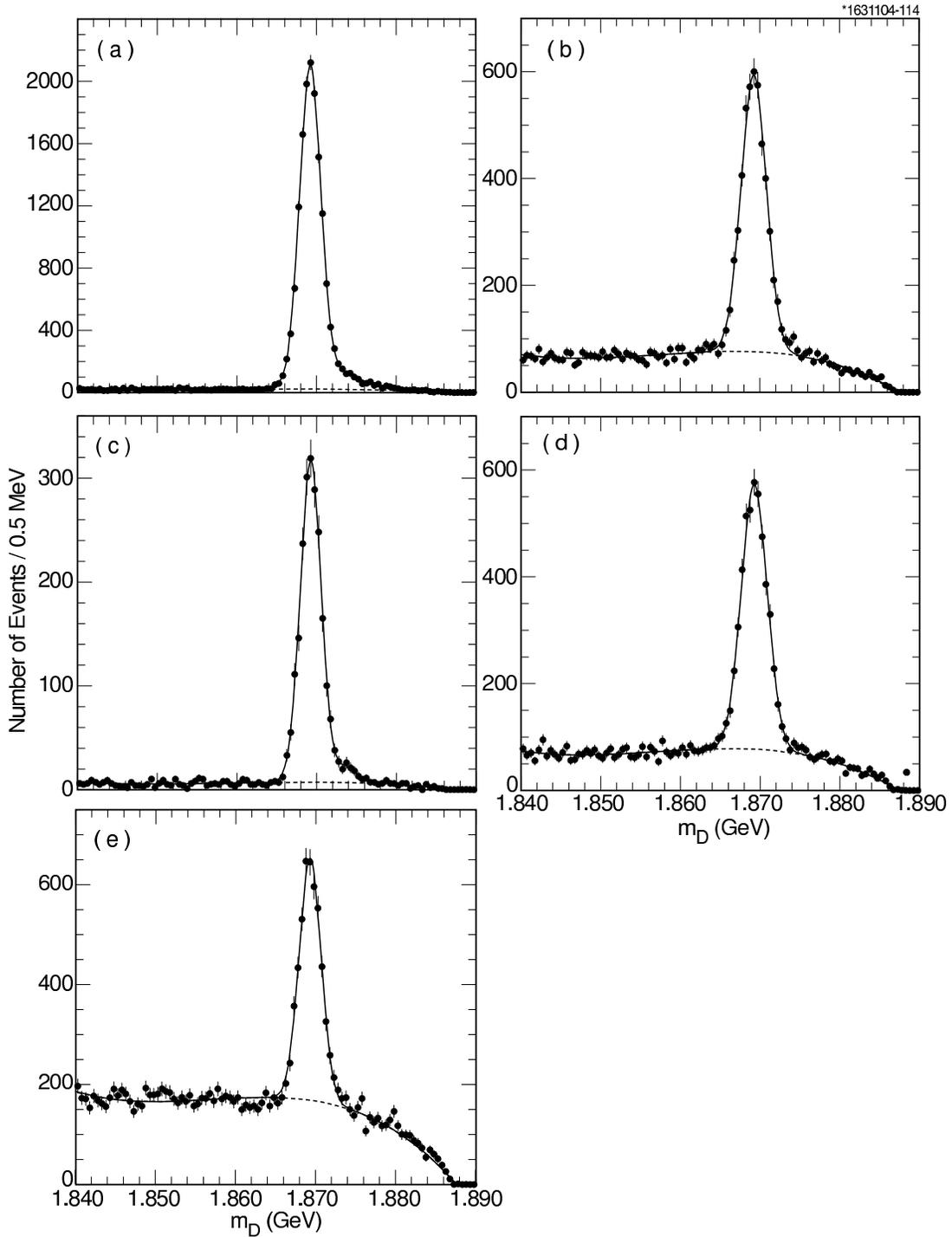} }
\caption{ Beam-constrained mass distributions for different fully
reconstructed $D^-$ decay candidates in the modes:
(a) $D^- \to K^+
\pi^- \pi^-$, (b) $D^-\to K^+ \pi^- \pi^- \pi^0$, (c) $D^- \to K_S\pi^-$, (d) $D^- \to K_S \pi^-\pi^-\pi^+$ and (e) $D^- \to
K_S\pi^- \pi^0$.
The solid curves show the sum of
Gaussian signal functions and $3^{rd}$ order polynomial background
functions. A single signal Gaussian is used for all modes except
for modes (a) and (c) where the sum of two Gaussians are used. The dashed
curves indicate the background fits.} \label{Drecon}
\end{figure}

The $m_D$ distributions for all $D^-$ tagging modes considered in
this data sample are shown in Fig.~\ref{Drecon} and listed in
Table~\ref{tab:Drecon} along with the numbers of signal events and
background events within $\pm$3 r.m.s. widths of the peak. The event numbers are determined from fits of
the $m_D$ distributions to Gaussian signal functions plus a
background shape. We fit with two different background
parametrizations: (a) a $3^{rd}$ order
polynomial, (b) a shape function
analogous to one first used by the ARGUS collaboration \cite{ARGUS} which has
approximately the correct threshold behavior at large $m_D$;
To use this function, we first fit it
to the data selected by using  $\Delta E$ sidebands, mode by mode,
defined as 40 MeV $<|\Delta E|<$ 60 MeV to fix the shape
parameters in each mode allowing the normalization
to float. For the $K^+\pi^-\pi^- \pi^o$, $K_S\pi^-\pi^-\pi^+ $ and
$K_S\pi^-\pi^o $ modes we use a single Gaussian to describe the
signal whose mass and width are allowed to float. For the
$K^+\pi^-\pi^- $ and $K_S\pi^-$ modes, where we see a small tail
on the higher mass side, we
use the sum of two Gaussian's for a signal function \cite{ISR}; in this
case both the means and widths of both Gaussians are allowed
to float.

\begin{table}[htb]
\begin{center}
\begin{tabular}{lcc}
    Mode  &  Signal           &  Background \\ \hline
$K^+\pi^-\pi^- $ & $15173 \pm 140$   & $~~583$\\
$K^+\pi^-\pi^- \pi^o$ & $4082 \pm 81$  & $1826$\\
$K_S\pi^-$ &   $2124\pm 52$& $~~251$\\
$K_S\pi^-\pi^-\pi^+ $ &  $3975 \pm 81$ & $1880$\\
$K_S\pi^-\pi^o $ &  $3297 \pm 87$ & $4226$\\
\hline
Sum &  $ 28651\pm207$ & $8765$\\
\hline\hline
\end{tabular}
\end{center}
\caption{Tagging modes and numbers of signal and background events
determined from the fits shown in Fig.~\ref{Drecon}.}
\label{tab:Drecon}
\end{table}
The difference between using the polynomial and ARGUS shapes in
the signal yields is  $\pm 2.2$\%, which we use as an estimate of the systematic error.

Selecting those candidates within 3 r.m.s. widths of the $D^-$
mass reduces the signal number by 77 events giving a total of
28,574$\pm$207$\pm$629 single tag events that we use for further analysis.
In the case of two Gaussians the wider width was used.

\section{${\boldmath D^+\to \mu^+\nu_{\mu}}$ Selection Criteria}
\label{sec:muonsel}
Using our sample of
$D^-$ event candidates we search for events with a single
additional charged track presumed to be a $\mu^+$. Then  we infer
the existence of the neutrino by requiring a measured value near
zero (the neutrino mass) of the missing mass squared (MM$^2$)
defined as
\begin{equation}
{\rm
MM}^2=\left(E_{beam}-E_{\mu^+}\right)^2-\left(-\overrightarrow{p}_{\!D^-}
-\overrightarrow{p}_{\!\mu^+}\right)^2, \label{eq:MMsq}
\end{equation}
where $\overrightarrow{p}_{D^-}$ is the three-momentum of the
fully reconstructed $D^-$.

We need to restrict the sample to candidate $\mu^+ \nu_{\mu}$
events resulting from the other $D$. Thus we wish to exclude
events with more than one additional track with opposite charged
to the tagged $D$, which we take
to be the muon candidate, or with extra neutral energy. It is
possible, in fact even likely, that the decay products of the
tagging $D^-$ interact in the detector material, mostly the EM
calorimeter and spray tracks and neutral energy back into the rest
of the detector. To evaluate the size of these contributions we
use a very pure sample of events obtained by finding fully
reconstructed $D^o\overline{D}^o$ events. The numbers of these
events in various decay modes are listed in
Table~\ref{tab:double}, a total of 782 events.

\begin{table}[htb]
\begin{center}
\begin{tabular}{llc}
    Mode 1 &  Mode 2           &  \# of events \\\hline
$K^-\pi^+$  & $K^+\pi^-$  & ~~89 \\
$K^+\pi^-\pi^+\pi^-$  & $K^-\pi^+$  &392 \\
$K^+\pi^-\pi^+\pi^-$ & $K^-\pi^+\pi^-\pi^+$  & 301 \\
\hline\hline
\end{tabular}
\end{center}
\caption{Fully reconstructed $D^o\overline{D}^o$ events }
\label{tab:double}
\end{table}

The number of interactions of particles with material and their
consequences depend on the number of particles, the kind of
particles and their momenta. Thus, the sum over these neutral $D$
decay modes isn't quite the same as the sum over the tagging $D^-$
decay; however, after accounting for the differences between the pion-nucleon and kaon-nucleon cross sections and the different momentum distributions of the tracks, we
find that the average over these modes is quite similar to the
$D^-$ tagging modes for this level of statistics.

Extra tracks do appear in these $D^o\overline{D}^o$ events. None
of these tracks, however, approach the main event vertex.
Requiring that good tracks are within 5 cm along the beam and
5 mm perpendicular to the beam does not include any additional
tracks from interactions in the material. We also reject $D^-$ tags with
additional $K_S \to \pi^+\pi^-$ candidates.

\begin{figure}[htbp]
\centerline{ \epsfxsize=3.0in \epsffile{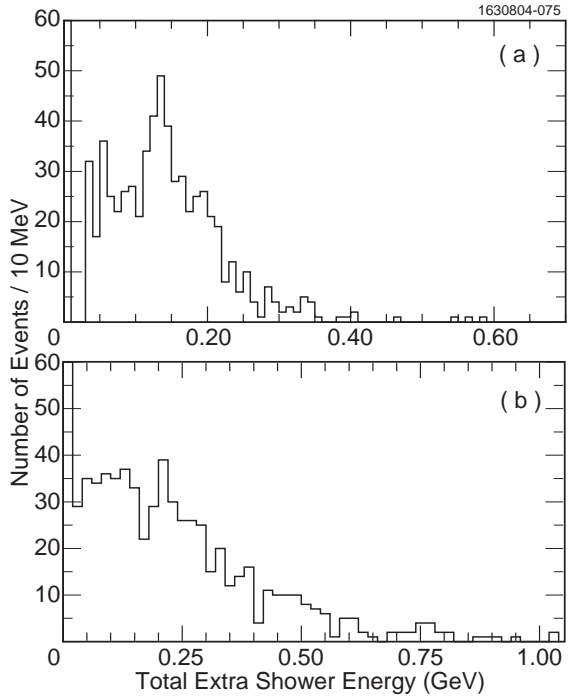} }
\caption{ Largest (a) and total extra shower (b) energies in the
$D^o\overline{D}^o$ sample. In both cases the first
bin is truncated; each plot has 782 total entries.} \label{Esh}
\end{figure}

In the $D^o\overline{D}^o$ events, energy in the calorimeter not
matched to any of the charged tracks is shown in Fig.~\ref{Esh}.
Figure~\ref{Esh}(a) shows the energy of the largest shower and
~\ref{Esh}(b) shows the total.
 We accept only as extra showers those that do not match a charged
 track within a connected region.  A connected region is a group of adjacent crystals
 with energy depositions which are nearest neighbors. This suppresses hadronic
 shower fragments which would otherwise show up as unmatched
 showers. Hadronic interactions and very
 energetic $\pi^o$'s tend to produce one connected region with many
 clusters.
 For further analysis we require that the largest unmatched shower
not to be larger than 250 MeV. This requirement is
(93.5$\pm$0.9)\% efficient for signal events, estimated from the
distribution of extra energies in the $D^o\overline{D}^o$ tag sample.
We assign an additional 4\% systematic error, due to the difference in our double tag and single tag samples.

\begin{figure}[htbp]
\centerline{ \epsfxsize=4.0in
\epsffile{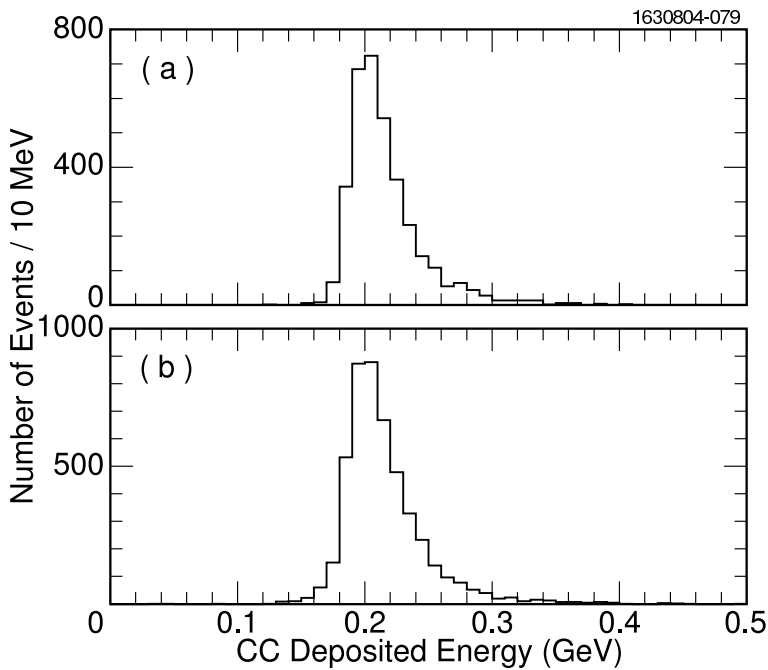} } \caption{
Deposited energy in the crystal calorimeter of muons created in the process
$e^+e^-\to \mu^+\mu^-$ from (a) data and (b) Monte Carlo.}
\label{mu-data-mupair}
\end{figure}

The muon candidate is required to be within the barrel region of
the detector $|\cos\theta|<0.81$; this requirement
insures that the MM$^2$ resolution is good as tracks at larger
angles cross fewer tracking layers and consequently are measured with poorer
precision.  In addition, this requirement helps reject background
from the decay $D^+\to\pi^+\pi^o$; this mode also gives a MM$^2$
near zero. Requiring the muon candidate in the barrel region (the
$\pi^+$ in this case) avoids having the photons from this decay
being lost in the transition region of the calorimeter between the
barrel and the endcap, because the $\pi^o$ direction is almost
directly opposite the $\pi^+$. Furthermore, the muon candidate is
required not to be consistent with the kaon hypothesis using RICH
information. Finally, we also require that the muon candidate
deposits less than 300 MeV of energy in the calorimeter,
characteristic of a minimum ionizing particle. This requirement is
very efficient for real muons, and rejects about 40\% of the pions
as determined using a sample of reconstructed $D^o\to K^-\pi^+$
decays. Figure~\ref{mu-data-mupair} shows the muon deposited energy
in the EM calorimeter both from data on $e^+e^-\to \mu^+\mu^-$ and
from GEANT simulation of the same process. The Monte Carlo and
data are in excellent agreement for muon shower energies.
We therefore use a GEANT
simulation of $D^+\to\mu^+\nu$ with lower energy muons to
determine that the efficiency of the calorimeter energy cut is
$(98.7 \pm 0.2)\%$.

When evaluating MM$^2$ using Eq.~(\ref{eq:MMsq})
there are two important considerations that are not obvious.
First of all, we explicitly need to take into account the crossing angle between the $e^+$ and $e^-$ beams.
This angle is about 4 mrad, varying slightly run to run; we use this information and Lorentz transform all
laboratory quantities to the center-of-mass. Secondly, we change the reconstructed
$D^-$ momenta so that they give exactly the known $D^-$ mass;
this changes and improves somewhat our knowledge of the $D^-$ direction.

\begin{figure}[htbp]
\centerline{ \epsfxsize=5.0in \epsffile{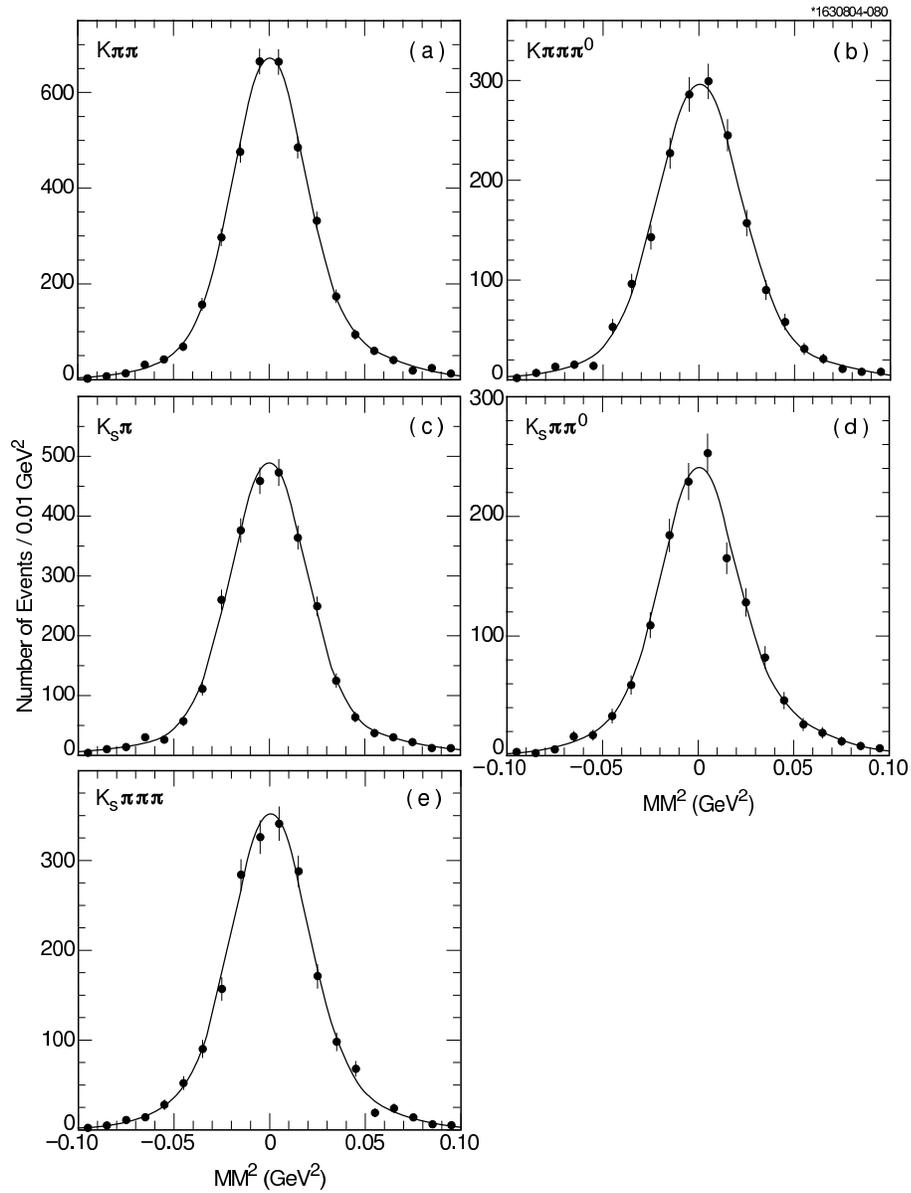} }
\caption{Monte Carlo simulation of $D^+\to \mu^+ \nu_{\mu}$ events for different
tags. The plots have been fitted to two Gaussians centered at
zero where the second Gaussian constitutes around 30\% of area.}
\label{mc-mm2}
\end{figure}

The MM$^2$ from Monte Carlo simulation is shown for our different tagging
samples in Fig.~\ref{mc-mm2}. The signal is fit to a sum of two Gaussians with
the wider Gaussian having about 30\% of the area independent of tagging mode.
The resolution ($\sigma$) is defined as
\begin{equation}
\sigma = f_1\sigma_1+(1-f_1)\sigma_2,
\end{equation}
where $\sigma_1$ and $\sigma_2$ are the individual widths of the
two Gaussians and $f_1$ is the fractional area of the first
Gaussian. The resolution is approximately 0.025 GeV$^2$ consistent
among all the tagging decay modes.

\begin{figure}[htbp]
\centerline{ \epsfxsize=3.0in \epsffile{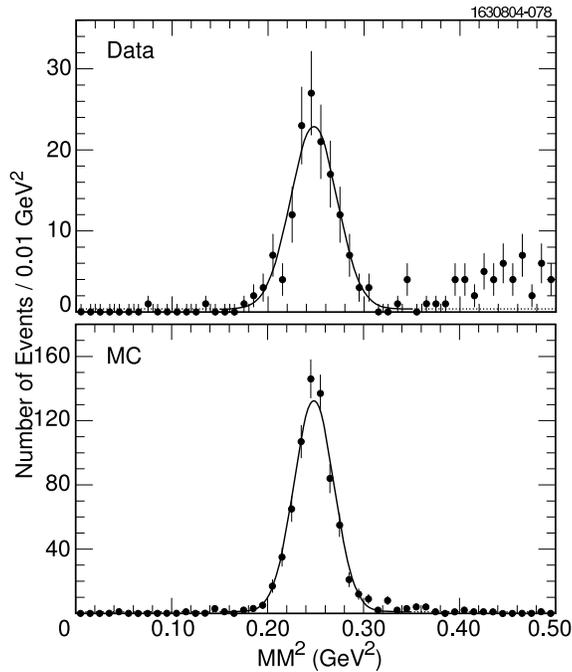} }
\caption{MM$^2$ distribution for the decay $D^+\to K_S\pi^+$ from
data and signal Monte-Carlo simulation} \label{mc-data-check}
\end{figure}

 We check our simulations using the $D^+\to K_S\pi^+$ decay. Here
 we choose events with the same requirements as used to search for
 $\mu^+\nu$ but require one additional found $K_S$. The MM$^2$
 distribution for this final state is shown in Fig.~\ref{mc-data-check}
 and peaks as expected at the $K_S$ mass-squared of 0.25 GeV$^2$. The
 resolution is measured to be 0.024$\pm$0.002 GeV$^2$ from a
 single Gaussian fit, consistent with but slightly larger than the Monte
Carlo estimate of 0.021$\pm$0.001 GeV$^2$. To account for the
difference in resolution between data and simulations we scale the
resolution by 14\% to 0.028 GeV$^2$ when looking for the
$D^+\to\mu^+\nu_{\mu}$ signal.

\begin{figure}[htbp]
\centerline{ \epsfxsize=3.0in \epsffile{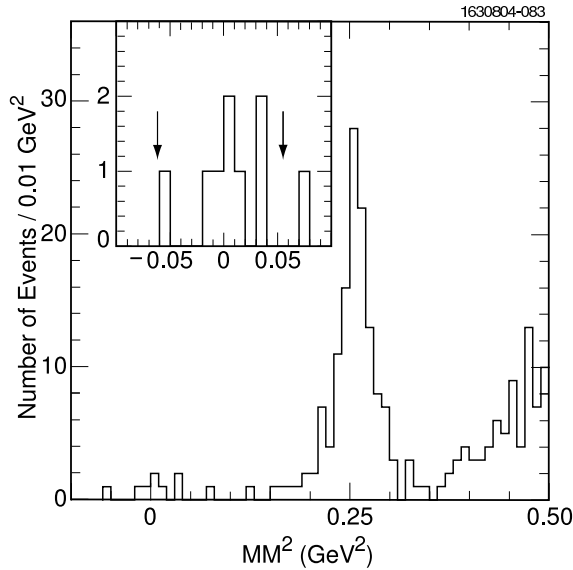} }
\caption{MM$^2$ using $D^-$ tags and one additional opposite sign
charged track and no extra energetic showers (see text). The
insert shows the signal region for $D^+\to\mu^+\nu$ enlarged; the $\pm2\sigma$
range is shown between the two arrows.}
\label{mm2}
\end{figure}

The MM$^2$ distributions for our tagged events requiring no extra
charged tracks besides the muon candidate and showers above 250
MeV as described above is shown in Fig.~\ref{mm2}.  We see a small
signal near zero containing 8 events within a 2$\sigma$ interval,
$-0.056$ GeV$^2$ to +0.056 GeV$^2$. This signal is most likely due
to the $D^+\to\mu^+\nu_{\mu}$ mode we are seeking. The large peak
centered near 0.25 GeV$^2$ is from the decay $D^+\to
\overline{K}^o\pi^+$ that is far from our signal region and is
expected since many $K_L$ would escape our detector.

\begin{figure}[htbp]
\centerline{ \epsfxsize=4.0in
\epsffile{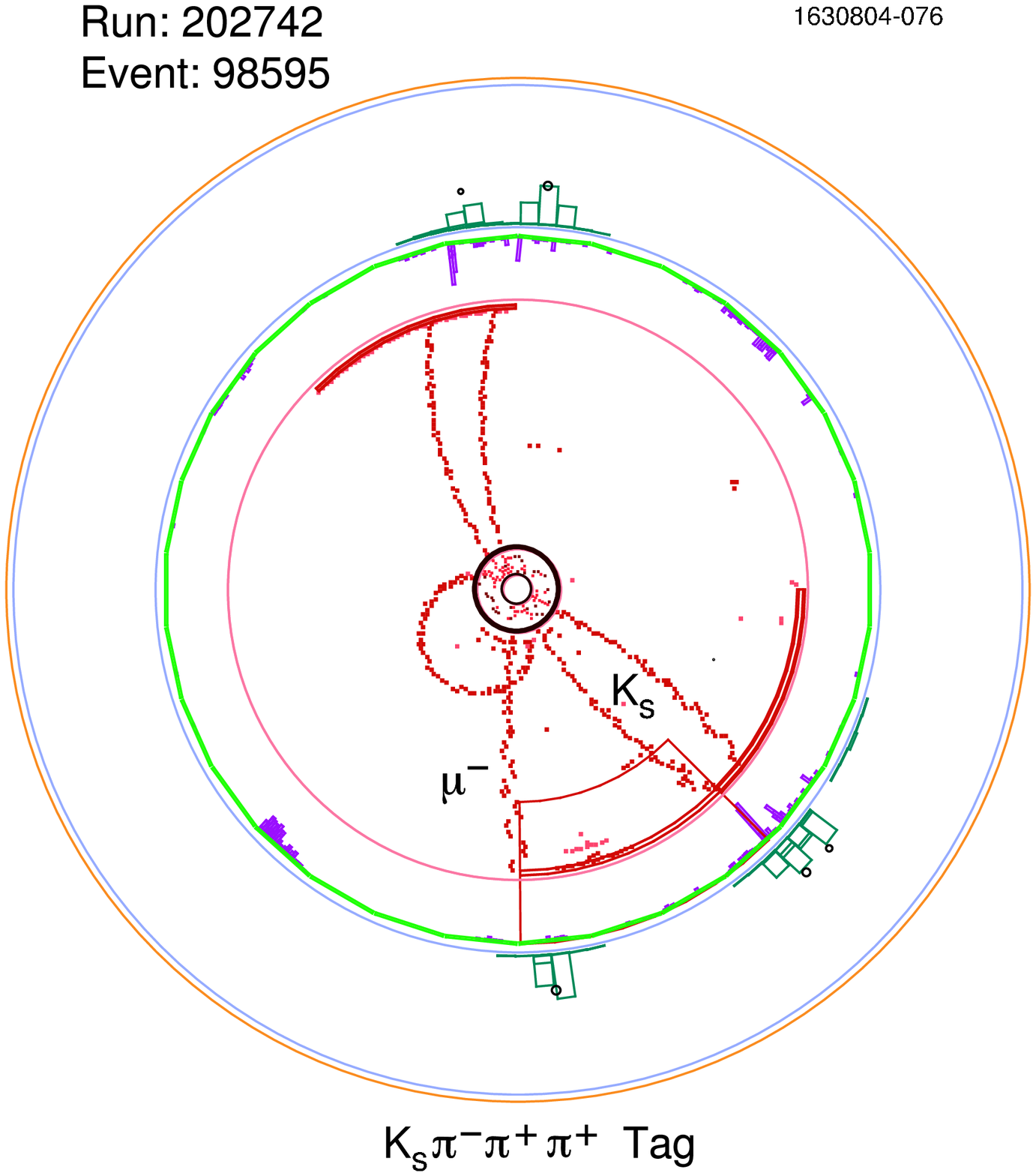} } \caption{A typical
$D^-\to\mu^- \overline{\nu}_{\mu}$ event. The tag is this case is
$D^+\to K_S\pi^-\pi^+\pi^+$. The muon and
the two oppositely charged pions forming the $K_S$ are indicated. The
$\pi^-$
is the ``curler" track with momentum around 50 MeV.}
\label{event}
\end{figure}

Table~\ref{tab:Muprop} lists the properties of each muon
candidate from the 8 events in the signal region.
A typical event is shown in Fig.~\ref{event}.

\begin{table}[htb]
\begin{center}
\begin{tabular}{lccrrc}
Tag&MM$^2$&CC energy  & $-2\log({\cal L}_K)$ &$-2\log({\cal L}_{\mu}$) &$\mu^{\pm}$\\
&  (GeV$^2$) & of $\mu^+$(GeV)  & &  &\\
\hline
 $K\pi\pi\pi^o$     & ~0.032  &0.186 &-4.3   &-166.0  &~+\\
 $K_S\pi$& ~-0.019  &0.201 &0.0    &-140.0       &~-\\
 $K\pi\pi$      &-0.051   &0.190 &~31.9    &-252.9 &~+\\
 $K\pi\pi$     &-0.004   &0.221 &0.0   &-115.2   &~+\\
 $K_S\pi\pi^o$     &0.032   &0.164 &~-0.3   &~-130.6 &~-\\
 $K_S\pi\pi\pi$&~0.001 &0.245 &-11.7   &-138.9 &~+\\
 $K\pi\pi\pi^o$& ~0.002  &0.204 &-8.6    &-88.6  &~-\\
 $K_S\pi\pi^o$  & ~0.014  &0.208 &-8.3    &-113.0     &~+\\

\hline\hline
\end{tabular}
\end{center}
\caption{Muon Candidate Properties. (CC indicates the crystal calorimeter.) }\label{tab:Muprop}
\end{table}


\section{Background Evaluation}
\subsection{Introduction}

\begin{figure}[htbp]
\centerline{ \epsfxsize=4.0in \epsffile{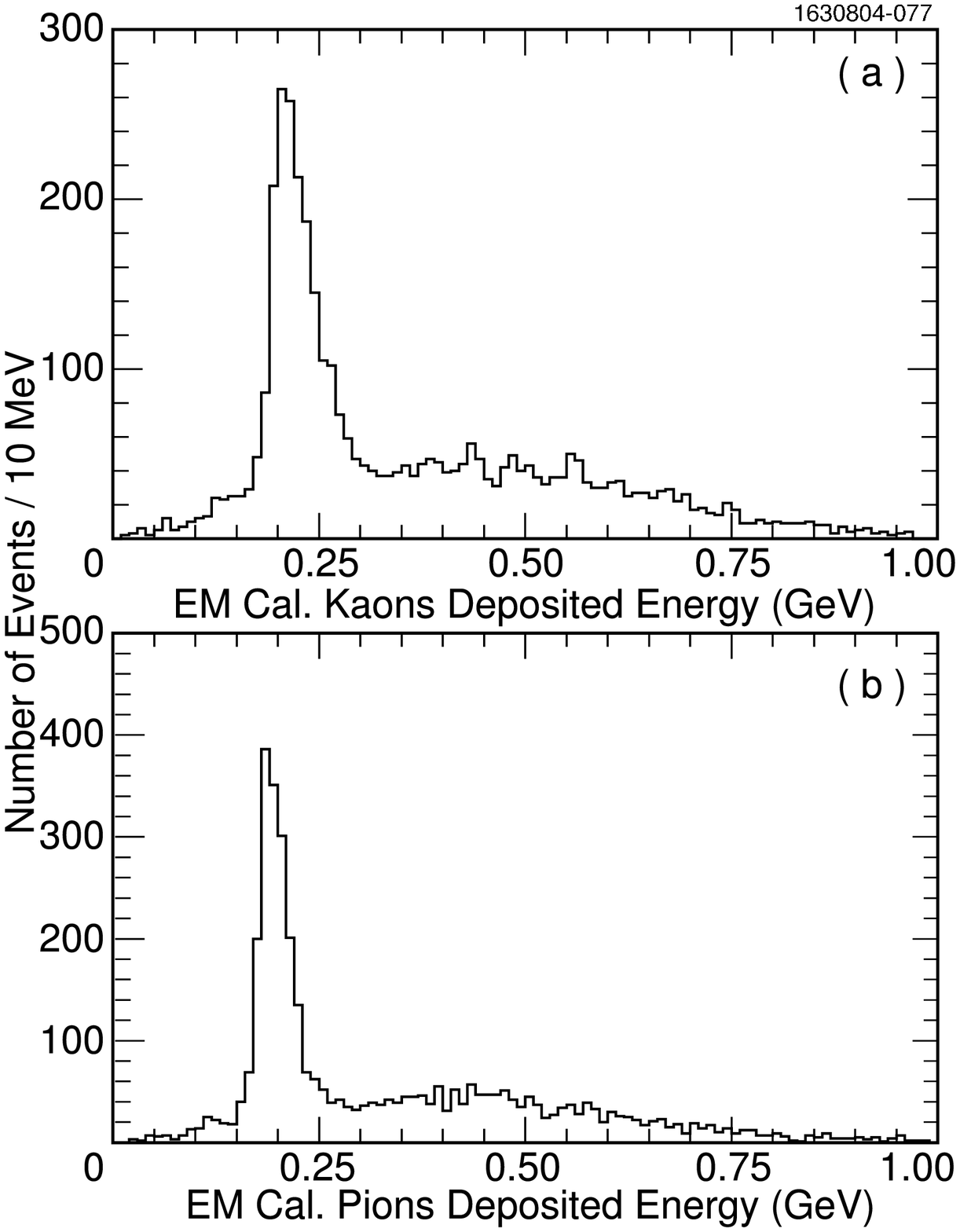} }
\caption{ Deposited energy in EM calorimeter for (a) kaons, (b)
pions from $D^0 \to K^-\pi^+$.} \label{kapi-dep}
\end{figure}

There are several background sources we need to evaluate. These
include background from other $D^+$ modes, background from
misidentified $D^o\overline{D}^o$ events and continuum background.
The requirement of the muon depositing $<$300 MeV in the
calorimeter, while about 99\% efficient on muons, rejects
only about 40\% of pions as determined from the
$D^o\overline{D}^o$ event sample where the pion from the
$K^{\pm}\pi^{\mp}$ mode was examined. In Fig.~\ref{kapi-dep} we
show the deposited energy in the calorimeter for both kaons and
pions obtained from the $K \pi$ tag sample.
\subsection{${\boldmath D^+}$ Backgrounds}

\begin{figure}[htbp]
\centerline{ \epsfxsize=4.0in \epsffile{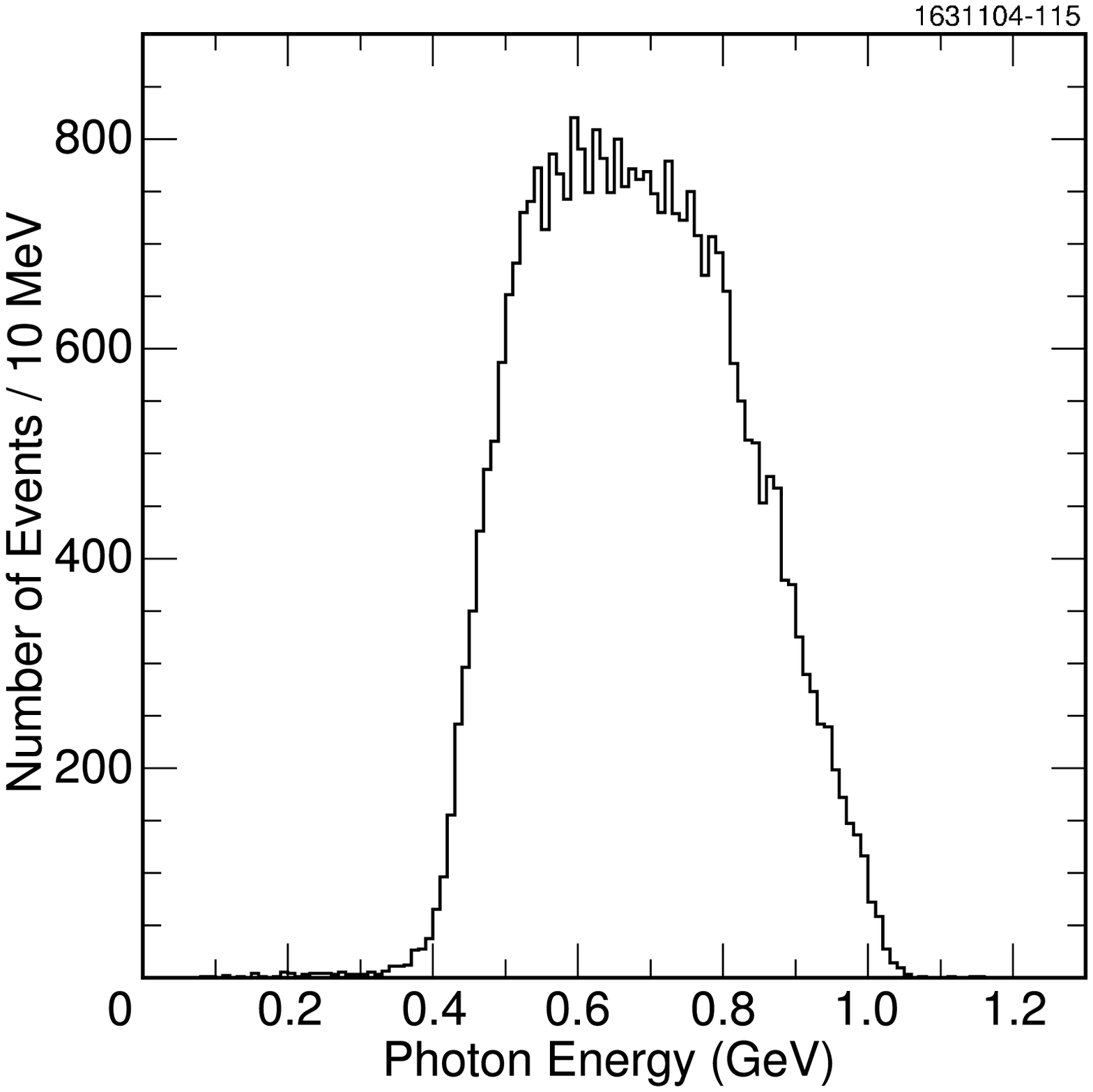} }
\caption{Maximum photon energy of the $\pi^o$ in the $D^+\to
\pi^+\pi^o$ decay from a GEANT simulation.} \label{mc-max}
\end{figure}

There are a few $D^+$ decay modes that could mimic the signal.
These are listed in Table~\ref{tab:Dpback} along with the
background estimate we obtained by Monte Carlo generation and
reconstruction of each specific mode. The branching ratios are
from the Particle Data Group except for the $\pi^+\pi^o$ mode
where a separate CLEO analysis gives a somewhat lower value
\cite{CLEOpipi}. This mode is the most difficult to reject because
the MM$^2$ peaks
 very close to zero, at 0.018 GeV$^2$, well within our resolution of 0.028 GeV$^2$. While we have insisted that
  the muon candidate be well within our acceptance, it is possible for the photons from the $\pi^o$ decay to
   inadvertently be matched to the tracks from the tagging $D^-$ or be
   missed. The maximum photon energy of the $\pi^o$ from a GEANT simulation of $D^+\to\pi^+\pi^o$ is shown in
   Fig.~\ref{mc-max}. We note that at least one photon from the
   $\pi^+\pi^o$ mode exceeds our 250 MeV calorimeter
   energy requirement and should in most cases cause such a decay
   to be vetoed.

\begin{figure}[htb]
\centerline{
\epsfxsize=4.0in \epsffile{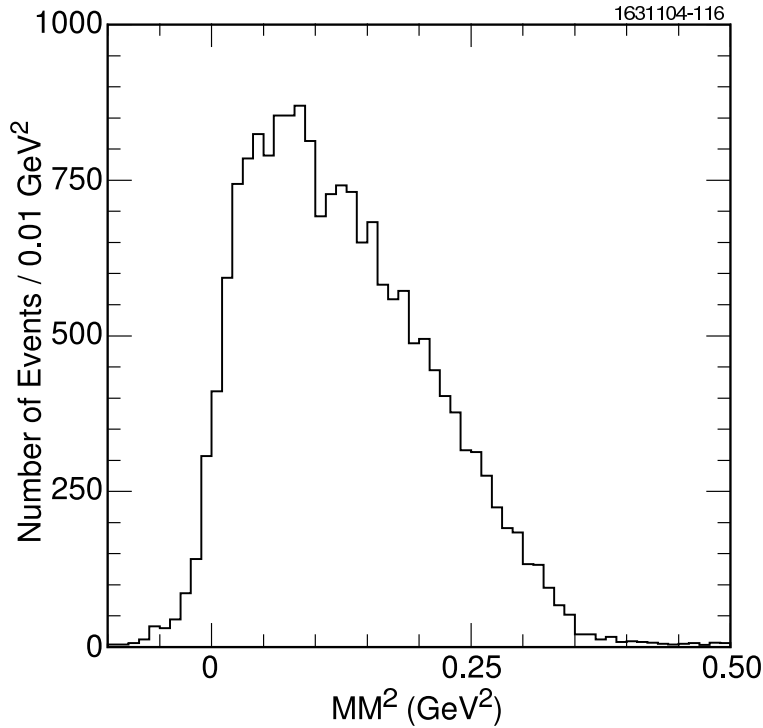}}
\caption{Missing
Mass squared distribution for $D^+ \to \tau^+\nu$ and $\tau^+\to
\pi^+\nu$.} \label{tau}
\end{figure}

   Even though the $\overline{K}^o\pi^+$ mode gives a large peak in the
   MM$^2$ spectrum near 0.25 GeV$^2$, our simulation shows that
   only a very small amount can enter our signal region, only 0.06
   events.
We have simulated backgrounds from $D^+\to\tau^+\nu$. Out of 10,000 simulated
events with $D^-$ tags, we found background only when $\tau^+\to \pi^+\nu$.
Because of
the small $D^+$-$\tau^+$ mass difference, the $\tau^+$ is almost at rest in the
laboratory frame and thus the $\pi^+$ has relatively large momentum causing the
MM$^2$ distribution to populate only the low MM$^2$ region, even in this case
with two missing neutrinos. The MM$^2$ distribution is
shown in Fig.~\ref{tau}.

\begin{table}[htb]
\begin{center}
\begin{tabular}{lll}
    Mode & ${\cal{B}}$ (\%) & \# of events \\\hline
$\pi^+\pi^o $ &0.13$\pm$0.02   & 0.31$\pm$0.04 \\
$\overline{K}^o\pi^+ $&  2.77$\pm$0.18  &0.06$\pm$0.05\\
$\tau^+\nu$& 2.64$\times$ ${\cal{B}}(D^+\to\mu^+\nu)$ & 0.30$\pm$0.07 \\
$\pi^o\mu^+\nu$& $0.25\pm 0.15$  & negligible \\\hline
sum & & 0.67$\pm$0.09\\
\hline\hline
\end{tabular}
\end{center}
\caption{Backgrounds from specific $D^+$ decay modes }
\label{tab:Dpback}
\end{table}

The semileptonic mode $\pi^o\mu^+\nu_{\mu}$ is similar to
$\pi^+\pi^o$ except that the $\pi^o$ often carries off enough
momentum to result in large MM$^2$. We found no candidate
background events in a Monte Carlo sample consisting of 50,000
tags plus a $D^+\to \pi^o\mu^+\nu$ decay.

\subsection{{$\boldmath D^o\overline{D}^o$} and Continuum Backgrounds}

These backgrounds are evaluated by analyzing Monte Carlo samples
corresponding to 5.2 times the total amount of data in our
possession. To normalize our Monte Carlo events to our data sample
we used $\sigma_{D^o\overline{D}^o}=3.5$ nb and
$\sigma_{continuum}=14.5$ nb \cite{sighad}. In each sample we found one
background event within two standard deviations of zero. These
correspond to 0.16$\pm$0.16 $D^o\overline{D}^o$ events and
0.17$\pm$0.17 continuum events forming background. As a check on the continuum background we analyzed 23 pb$^{-1}$ of continuum data taken a center-of-mass energy of 3670 MeV. We didn't find any $D^+\to\mu^+\nu$ candidate events.

\subsection{Background Summary}

Our total background is 1.00$\pm$0.25 events. The probability of one background event fluctuating to 8 or more signal events is only $10^{-5}$, and even including the 0.25 event uncertainty in the background the signal has greater than five standard deviation significance.
Because of
the uncertainties in the Monte Carlo simulation we assign a 100\%
error to our background estimate: 1.0$\pm$1.0 events, for the purpose
of evaluating the branching ratio.

\section{Branching Ratio and Decay Constant}

Subtracting the 1.0 event background from our 8 events in the
signal region, we determine a branching fraction using a detection
efficiency for the single muon of 69.9\%. This efficiency includes
the selection on MM$^2$ within $\pm2\sigma$ limits,
the tracking, the particle identification, the probability of the
crystal energy being less than 300 MeV,  and the probability of not
having another unmatched shower in the event with energy greater
than 250 MeV. We assign a relative 5.3\% error on this efficiency,
the components of which are shown in Table~\ref{tab:eff}. We use a
$3\%$ systematic error on track finding found using the double
tagged events and we estimate the error on the particle
identification cut to be $1\%$ from studies of $D^{*+}$ decays in
higher beam energy data. The error on the minimum ionization cut on the muon candidate in the calorimeter is 0.2\% and discussed in detail in section~\ref{sec:muonsel}. A 4\% dominantly systematic error due to rejection
of events with excess shower energy is assigned to the efficiency of this
cut determined
by using the $D^o\overline{D}^o$
sample and also discussed in section~\ref{sec:muonsel}.
\begin{table}[htb]
\begin{center}
\begin{tabular}{lc}
     & Systematic error (\%) \\ \hline
MC statistics &0.8  \\
Track finding &3 \\
PID cut &1 \\
Minimum ionization cut &1 \\
Extra showers cut &4  \\\hline
Total &5.3\\
 \hline\hline
\end{tabular}
\end{center}
\caption{Systematic errors on the $D^+ \to \mu^+ \nu_{\mu}$
efficiency.} \label{tab:eff}
\end{table}

To compute the branching ratio we use 7.0$\pm$2.8 signal events
divided by 69.9\% and the 28574 $D^{\mp}$ tags. No other
efficiencies enter.
The systematic error on the branching fraction arises from the 5.3\% systematic error on the efficiency, a 2.2\% systematic error in the number of $D^-$ tags and a 15.4\% systematic error on the background. The total systematic error, evaluated by adding these contribution in quadrature, is 16.4\%.
Our result for the branching fraction is
\begin{equation}
{\cal{B}}(D^+\to\mu^+\nu_{\mu})=(3.5\pm 1.4 \pm 0.6)\times
10^{-4}~.
\end{equation}

The decay constant $f_{D^+}$ is then obtained from Eq.~(\ref{eq:equ_rate}) using 1.04 ps as the $D^+$ lifetime and 0.224 as $|V_{cd}|$ \cite{PDG}.
Our final result is
\begin{equation}
f_{D^+}=(202\pm 41\pm 17)~{\rm MeV}~.
\end{equation}

\section{Conclusions}

There have been several experimental studies of $D$ meson decay
constants. The Mark III group published an upper limit of
${\cal{B}}(D^+\to\mu^+\nu_{\mu})<7.2 \times 10^{-4}$, which leads
to an upper limit on the decay constant  $f_{D^+}<290$ MeV at 90\%
confidence level based on 9.3 pb$^{-1}$ of data taken on the
$\psi''$ \cite{MarkIII}. BES claimed the observation of one event
at a center-of-mass energy of 4.03 GeV with a branching ratio of
$(0.08^{+0.17}_{-0.05})$\% \cite{Rong1}. Recently, using 33
pb$^{-1}$ of $\psi''$ data they presented 3 event candidates and
with an estimated background of 0.33 events where neither
$\pi^+\pi^o$, or $\tau^+\nu$ were mentioned as a possible
background modes, nor was continuum background considered
\cite{Rong2}. Here they find a branching ratio of
$(0.122^{+0.111}_{-0.053} \pm 0.010)$\%, and a corresponding value
of $f_{D^+}=(371^{+129}_{-119} \pm 25)$ MeV. Our value is
considerably smaller, though compatible with their large error.

Our analysis shows the first statistically significant signal for
$D^+\to\mu^+\nu$. The branching fraction is
\begin{equation}
{\cal{B}}(D^+\to\mu^+\nu)=(3.5\pm 1.4 \pm 0.6)\times 10^{-4}~,
\end{equation}
and the decay constant is
\begin{equation}
f_{D^+}=(202\pm 41\pm 17)~{\rm MeV}~.
\end{equation}

Our result for $f_{D^+}$, at the current level of precision, is
consistent with predictions of lattice QCD and models
listed in
Table~\ref{tab:Models}.

\begin{table}[htb]
\begin{center}
\begin{tabular}{lcl}
    Model  &  $f_{D^+}$ (MeV)          &  $f_{D_s^+}/f_{D^+}$           \\\hline
Lattice QCD (Fermilab and MILC) \cite{Lat:Milc} &
$225^{+11}_{-13} \pm 21 $&$1.17\pm 0.06\pm 0.06$ \\
Quenched Lattice QCD (UKQCD) \cite{Lat:UKQCD} & $210\pm 10^{+17}_{-16}$ & $1.13\pm 0.02^{+0.04}_{-0.02}$\\
QCD Spectral Sum Rules \cite{Chiral} & $203\pm 20$ & $1.15\pm 0.04$ \\
QCD Sum Rules \cite{Sumrules} & $195\pm 20$ & \\
Relativistic Quark Model \cite{Quarkmodel} & $243\pm 25$ & 1.10 \\
Potential Model \cite{Equations} & 238  & 1.01 \\
Isospin Mass Splittings \cite{Isospin} & $262\pm 29$ & \\
\hline\hline
\end{tabular}
\end{center}
\caption{Theoretical predictions of $f_{D^+}$ and
$f_{D_s^+}/f_{D^+}$} \label{tab:Models}
\end{table}

The models generally predict $f_{D_s^+}$ to be 10-15\% larger than
$f_{D^+}$. CLEO previously measured $f_{D_s^+}$ as ($280\pm 19\pm
28\pm 34)$ MeV \cite{chadha}, and we are consistent with these
predictions as well. We look forward to more data to improve the
precision.
\section{acknowledgments}
We gratefully acknowledge the effort of the CESR staff in
providing us with excellent luminosity and running conditions.
This work was supported by the National Science Foundation and the
U.S. Department of Energy.


\begin{thebibliography}{99}
\bibitem{ISR}
The tail appears to be caused by initial state radiation.
\end{thebibliography}
\end{document}